\documentclass[11pt]{article}
\usepackage{epsfig,citesort}
\usepackage{multicol}
\begin{document}

\title{Heterogeneous nucleation near a metastable
vapour-liquid transition: the effect of wetting transitions}
\author{{\bf Richard P. Sear}\\
Department of Physics, University of Surrey\\
Guildford, Surrey GU2 7XH, United Kingdom\\
email: r.sear@surrey.ac.uk}
\maketitle

\begin{abstract}
Phase transformations such as freezing typically start with
heterogeneous nucleation.
Heterogeneous nucleation near a wetting transition, of a crystalline
phase is studied. The wetting transition occurs at or near a
vapour-liquid transition which occurs in a metastable fluid.
The fluid is metastable with respect to crystallisation, and
it is the crystallisation of this fluid phase that we are interested in.
At a wetting transition a thick layer
of a liquid phase forms at a surface in contact with
the vapour phase. 
The crystalline nucleus is then
immersed in this liquid layer, which
reduces the free energy barrier to nucleation and so dramatically
increases the nucleation rate. The variation in the rate of
heterogeneous nucleation close to wetting transitions
is calculated for systems in which the longest-range forces
are dispersion forces.
\end{abstract}

\section{Introduction}

When water is cooled below $0^{\circ}$C at atmospheric pressure
it freezes, it turns into ice. This conversion of one phase, water,
into another, ice, starts with the nucleation of a microscopic
nucleus of ice. This nucleus consists of only of order 10 molecules,
its formation costs free energy and occurs not in the bulk of the
water but at a surface in contact with the water.
The free energy cost provides a barrier to the nucleation of ice.
If the free energy cost or barrier is large it will limit the
rate of crystallisation.
When the barrier is
very large the phase which is not the equilibrium one, for example
water below $0^{\circ}$C, will persist for very long times.
The fluid phase is then called metastable. So to determine whether a phase
which is not an equilibrium phase is metastable or whether the
equilibrium phase nucleates rapidly we need to calculate the
free energy barrier to nucleation.
We do this here near to
and at {\em another}
phase transition, a phase transition between two phases neither of which
is the true equilibrium phase.
The process of nucleation at a surface is called heterogeneous nucleation
to distinguish it from
homogeneous nucleation which occurs in the bulk.
See Ref.~\cite{debenedetti} for an introduction to nucleation.

So, here we study the rate of heterogeneous nucleation of one
phase transition, a fluid-crystal phase transition, near a second
phase transition --- a phase transition between phases which are
both meta- or unstable with respect to crystallisation.
This second phase transition is a vapour-liquid transition.
At phase transitions the thermodynamic functions,
including interfacial tensions, exhibit singular behaviour which
is universal in the sense that many different systems show behaviour
which is identical up to a few scale factors. Here we show that
at the vapour-liquid transition the
free energy barrier to heterogeneous nucleation of a crystal shows behaviour
which although not truly
universal is the same near all vapour-liquid transitions,
up to a few scale factors, assuming that the longest-range
interactions are dispersion forces.
This is
due to wetting: the formation of a thick layer of the liquid phase
on a surface in contact with the vapour phase
\cite{cahn77,schick,indekeu,bonn01}.
The layer forms
as coexistence is approached and causes a 
drop in the nucleation barrier to the nucleation of
a dense phase such as a crystalline phase.
The dependence on the nature
of longest-range forces makes our findings not-quite universal
(unlike homogeneous nucleation near a bulk critical point \cite{searxxx}
which is universal).

\begin{figure}
\begin{center}
\epsfig{file=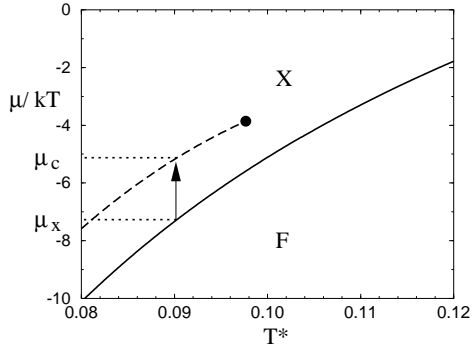,width=2.6in}
\caption{
\lineskip 2pt
\lineskiplimit 2pt
Bulk equilibrium phase diagram in the chemical potential--temperature
($\mu$-$T$) plane. The model is a simple model of a globular protein which
has a metastable fluid-fluid transition, denoted by the dashed curve,
which ends at a critical point, the black dot.
The region in the chemical potential-temperature plane where
the equilibrium phase is the crystalline (fluid) phase is denoted by an X (F).
For one particular temperature the
chemical potential of the fluid-crystal transition, $\mu_x$,
and that at metastable fluid-fluid coexistence,
$\mu_c$ are marked on the diagram.
The reduced
temperature $T^*=kT/\epsilon$ where $\epsilon$ is a bond energy.
\label{mut}
}
\end{center}
\end{figure}

Motivation for our study is provided by the fact that some globular
proteins have the correct set of phase transitions to observe
heterogeneous nucleation of a crystal near a vapour-liquid transition.
See Refs.~\cite{broide91,muschol97} for the phase diagrams of
a number of globular proteins.
The crystallisation of globular proteins
is of great interest because protein crystals are required in
order to determine the all-important three-dimensional structure
of a protein \cite{rosenberger96}. Also, although we will always
refer to the nucleus as being crystalline and the other transition
as being a vapour-liquid transition, our findings are much
more general. They refer to the nucleation of any noncritical phase
near another, Ising-type, phase transition. We simply describe the
phases as crystal, vapour and liquid for simplicity and because
having definite phases in mind is useful for pedagogical purposes.

Earlier work has found
universal behaviour of the nucleation barrier for
{\em homogeneous} nucleation near a bulk critical point, see
Refs.~\cite{tenwolde97,talanquer98,sear01a,sear01b,searxxx}.
This earlier work, in particular that of ten Wolde and Frenkel
\cite{tenwolde97},
inspired this study of heterogeneous nucleation and the results
of Refs.~\cite{tenwolde97,talanquer98,sear01a,sear01b,searxxx,zohreh}
are in a sense the homogeneous nucleation analogues of the
results we will obtain for heterogeneous nucleation.
Also, Talanquer and Oxtoby have studied heterogeneous nucleation
of a liquid from a vapour phase \cite{talanquer96}. They studied the
nucleation of the liquid phase at a surface when the
liquid phase itself is close to wetting this surface. So, although
they studied heterogeneous nucleation in the vicinity of a wetting
transition, as we do below, they studied the nucleation of the liquid
phase, the phase which is doing the wetting, whereas here we study
the heterogeneous nucleation of another phase, the crystalline phase.

In the next section, we introduce both our model of the
process of heterogeneous nucleation and the phase behaviour
of the systems we are interested in.
In section \ref{wetnuc} we
derive the variation in the rate of
heterogeneous nucleation near wetting transitions.
Then in section \ref{homo}
we compare with homogeneous nucleation of a crystal near
a bulk critical point. The last section is a conclusion.

\section{Heterogeneous nucleation}
\label{hetero}

Heterogeneous nucleation is an activated process \cite{debenedetti,xtalwet}
and as such occurs at a rate which decreases exponentially with the
height of the barrier $\Delta F$, which must be overcome.
If $N_n$ is the number of nuclei per unit area
crossing the barrier per unit time then
$N_n$ is given by an expression of the form \cite{debenedetti}
\begin{equation}
N_n=\sigma\tau^{-1}\exp(-\Delta F/kT),
\label{rate}
\end{equation}
where $\sigma$ is a surface density, i.e., it has dimensions of
inverse area, and $\tau$ is a characteristic time.
We will refer to $N_n$ as the nucleation rate or heterogeneous
nucleation rate. The surface is smooth, perfectly planar
and chemically homogeneous.
Equation (\ref{rate}) comes from the fact that the nucleus is a large,
i.e., improbable, fluctuation. As a fluctuation its probability
of occurring in unit area is $\sigma\exp(-\Delta F/kT)$.  The rate
at which these fluctuations cross the barrier is then estimated as the
number of fluctuations divided by $\tau$ which is an estimate of how
long it takes the nucleus to acquire one or a few extra molecules,
enough for the nucleus to be big enough to grow irreversibly into
a crystallite. Equation (\ref{rate}) is far from rigorous but has been
found for homogeneous nucleation to be a reasonable estimate. See
the book of Debenedetti \cite{debenedetti} for a discussion
and \cite{wolde96} for a detailed comparison  of an expression of the
form of Eq.~(\ref{rate}) with the results of
computer simulation (for homogeneous nucleation).
For the remainder of this work we will
assume that
$\tau$ and $\sigma$ vary weakly with temperature and chemical potential
and so the variation of the rate of heterogeneous nucleation is
dominated by the variation in the free energy barrier $\Delta F$.
Work on homogeneous nucleation has shown this assumption to be most
often justified, except near a glass transition where $\tau$ increases
sharply. However, verifying it requires a detailed calculation for
a specific model system, which we do not do here.

Consider the phase diagram in the chemical-potential--temperature plane
shown in Fig.~\ref{mut}.
It is the phase diagram of a simple model of a globular protein,
calculated using an approximate theory.
See Ref.~\cite{sear99} for the precise definition
of the model; the model parameters
have the same values as they do for Fig.~4 of that reference.
Also, see this reference for the same phase diagram in the
density-temperature plane which can be seen to be qualitatively like
that of a number of globular proteins \cite{broide91,muschol97}.
At true equilibrium there is only one phase transition, from a fluid
phase, the equilibrium phase below the solid curve, to a crystal,
the equilibrium phase above the curve. However, if the barrier to
formation of the crystal phase is high then the chemical potential
can be increased at constant temperature along a path such as that
indicated by the arrow in Fig.~\ref{mut},
until the metastable fluid undergoes another
phase transition: a transition from a vapour phase to a liquid phase.
The vapour is below the dashed curve, the liquid above.
We have used the phase diagram of a model protein because it has
the correct form: it has
a vapour-liquid transition near where we expect the nucleation
of the crystalline phase to occur. However, the behaviour we find will
apply 
whenever heterogeneous nucleation occurs at a surface which passes
through a wetting transition.

\begin{figure}
\begin{center}
\epsfig{file=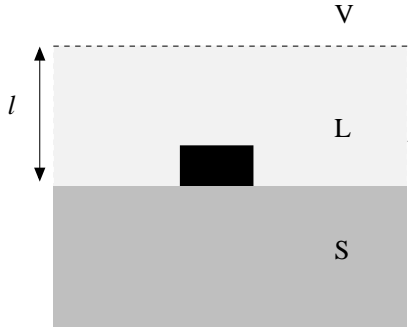,width=2.1in}
\end{center}
\caption{{
A schematic of a wetting layer atop a surface, with a nucleus of the
crystalline phase in contact with the surface and so within the
wetting layer. The material of the surface (S) is shaded dark gray, the
liquid (L) is lightly shaded and the nucleus is black. The vapour (V)
is left unshaded.
The thickness $l$ of the layer of liquid is indicated.
\label{fighet}
}}
\end{figure}

Figure \ref{fighet} is a schematic of a crystalline nucleus in
contact with a surface and immersed in a liquid layer of thickness $l$,
with a vapour phase on top. This is the situation of interest, there
is a bulk vapour phase against a smooth surface which attracts the
molecules causing a layer of liquid to form near and at vapour-liquid
coexistence.
At coexistence and if we are above the wetting temperature, see
Refs.~\cite{schick,bonn01} and section \ref{wetnuc}, there is
a very thick layer of liquid covering the surface and
separating the vapour phase from the
surface. The thickness $l$ is then limited only by gravity.
This wetting
layer will reduce the nucleation barrier greatly if the interfacial
tension between the liquid and the nucleus, $\gamma_{xl}$
is lower than that between the vapour and the nucleus, $\gamma_{xv}$.
If the surface area of the nucleus not in contact with the surface
is $S$, then the free energy reduction will be $S(\gamma_{xl}-\gamma_{xv})$
when the wetting layer forms. For $S$ of order 10 times the area
per molecule of the surface of a crystal, and the difference
$\gamma_{xl}-\gamma_{xv}$ of order $kT$ divided
by the area of one molecule, the reduction
in the surface contribution to the free energy barrier is of order
$10kT$. This reduction will occur on moving in the vapour phase
from conditions of chemical potential and temperature far
from vapour-liquid coexistence, where there is no wetting layer,
to at or very near coexistence.
A large reduction which will lead, Eq.~(\ref{rate}),
to a very large increase in the nucleation rate $N_N$, which
should be easily large enough
to observe in an experiment.

\section{Variation in the rate of heterogeneous
nucleation as a wetting transition is approached}
\label{wetnuc}

In this section we start with the assumption that the variation of the
rate is dominated by that in the free-energy barrier $\Delta F$ and
then calculate how $\Delta F$ varies near wetting transitions of different
types. Near a wetting transition the qualitative behaviour,
in particular the form of the singularities, can be determined
without knowing any specific
details of the interactions or of the phase which is nucleating,
we only require that the longest range interactions be dispersion forces
\cite{israelachvili}. This is true as the singularities come from
long lengthscale phenomena for which the small lengthscale chemical
details are irrelevant.

There are a number of different wetting
phase transitions, see for example the excellent review of Schick
\cite{schick}. We will deal with the three most common, starting with complete
wetting. In each case we will work very close to the wetting
transition, temperatures or chemical potentials very close to their
values at the transition. We will determine the leading order singular terms
in the temperature or chemical potential variation of the
nucleation rate near the transition.

\begin{figure}
\begin{center}
\epsfig{file=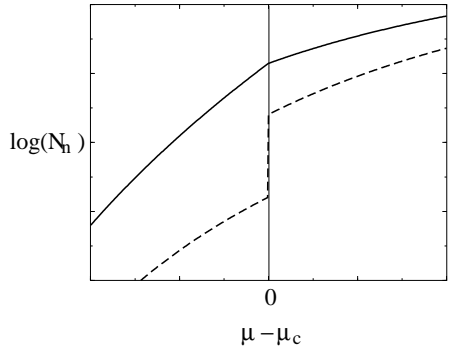,width=2.5in}
\end{center}
\caption{{
A schematic of the variation in nucleation rate with chemical potential
$\mu$ near vapour-liquid coexistence at $\mu_c$. For $\mu-\mu_c<0$ the
nucleation is occuring at a surface in contact with the vapour
and for $\mu-\mu_c>0$ the nucleation is occuring at a surface
in contact with the liquid phase. The solid curve is when there
is complete wetting of the surface by the liquid, the dashed curve for
when there is partial wetting.
\label{figcw}
}}
\end{figure}

\subsection{Complete wetting}

As heterogeneous nucleation occurs at a surface in contact with
the fluid it is rather obviously controlled by what happens at the
surface. Now, if the surface attracts the molecules of the fluid then
not-too-far from the critical point \cite{cahn77,indekeu,bonn01}, we will
have wetting. Wetting is where at vapour-liquid coexistence a thick
layer of the liquid phase interposes itself between the surface
and the vapour phase, replacing the surface-vapour interface
by a surface-liquid interface plus
a liquid-vapour interface. It is also called complete wetting \cite{schick}.
At coexistence the thickness of the layer
is generally limited only by gravity. Thus, our first result is the,
rather obvious, fact that if the surface is wet by a thick layer of
liquid the rate of heterogeneous nucleation will be the {\em same}
at surfaces in contact with the two coexisting phases, the vapour
and the liquid.
This result
is universal, it is true whenever the surface is wet.
The free energy
cost of forming a nucleus at the surface in contact with the bulk
vapour is the same as the cost of forming a nucleus at the surface
in contact with the bulk liquid, as both surfaces are covered by liquid.

This is at coexistence, the chemical potential $\mu=\mu_c$, where
$\mu_c$ is the chemical potential at vapour-liquid coexistence at
this temperature. In the vapour phase off coexistence, $\mu<\mu_c$,
there is still a layer of liquid at the surface provided $h=\mu-\mu_c$
is not too negative. But this layer thins as $h$ becomes more negative
as we move farther from coexistence. How it thins depends on the
nature of the longest-range forces present in the system \cite{schick}.
Here we assume that the longest range forces are dispersion forces,
which is most often the case. Then the free energy per unit area,
$f$, of a film of liquid
between the surface and the vapour of thickness $l$ is
\cite{israelachvili,schick}
\begin{equation}
f=\frac{a}{l^2}-h\delta\rho l,
\label{fcomp}
\end{equation}
where we have included only those parts which depend on the thickness
of the layer.
The first term is the contribution of the dispersion forces,
$a$ is a positive constant and the $l^{-2}$ dependence comes from starting with
the usual $l^{-6}$ dependence, integrating over a volume, for the solid
substrate,
and then over the thickness of the layer of liquid.
This fourfold integration changes the
$l^{-6}$ dependence to $l^{-2}$ \cite{israelachvili}.
The second term is the
increase in free energy due to the formation of a layer of liquid
of thickness $l$ when the chemical potential of the liquid, $\mu_c$,
is higher than the chemical potential. In this term $\delta\rho$ is
the difference in number density between the liquid and vapour phases.
Minimising the free energy
Eq.~(\ref{fcomp}) we obtain the thickness
\begin{equation}
l=\left(2a/(-h\delta\rho)\right)^{1/3},
\label{lcomp}
\end{equation}
which diverges at coexistence: this divergence is in practice
cutoff at some large thickness by gravity. The divergence at
coexistence is a genuine phase transition: $l^{-1}$ is analogous to an
order parameter and the exponent $1/3$ in Eq.~(\ref{lcomp})
is a critical exponent \cite{schick}.

So, what is the free energy barrier to heterogeneous nucleation
off coexistence: $h<0$ but small? We know that at coexistence the
liquid layer is very thick and so the free energy of a nucleus
in the vapour phase is the same as that in the liquid, call it
$\Delta F_L(\mu,T)$. As we move off coexistence the liquid layer thins and
the nucleus will notice this because it will interact with the vapour phase
once $l$ is not too large.
See Fig.~\ref{fighet} for a schematic of a nucleus at a substrate
in a liquid layer.
For $l$ larger than the radius of the nucleus,
the nucleus will interact with the vapour as a point object. The interaction
of a small object with the vapour across the liquid layer
varies as $l^{-3}$ and is
proportional to the volume of the nucleus, $v_n$ \cite{israelachvili}.
So the nucleus-vapour interaction increases the
nucleation barrier to $\Delta F$, where $\Delta F$ is given by
\begin{equation}
\Delta F(\mu,T)=\Delta F_L(\mu,T)+\frac{Av_n}{l^3} ~~~~~l~\mbox{large},
\label{fint}
\end{equation}
where $A$ is a coefficient for the interaction of the nucleus with the
vapour across a slab of liquid.
A dense nucleus
will generally repel a dilute vapour and
so $A$ will then be positive \cite{israelachvili,indekeu}.
Here, we focus on nucleation of a crystal phase and crystalline
phases are rather dense.
Also, note that $\Delta F_L(\mu, T)$ is a function of $\mu$ and we
will use it for $\mu<\mu_c$ where the vapour phase is more stable than
the liquid. We assume that we can continue $\Delta F_L(\mu,T)$ into 
the region where the liquid is metastable with respect to the
vapour,
$h$ negative but small, and that $\Delta F_L(\mu,T)$ is analytic
at $h=0$, at coexistence.
Using Eq.~(\ref{lcomp}) for the thickness we obtain
\begin{equation}
\Delta F(\mu,T)=\Delta F_L(\mu,T)-
h\frac{A\delta\rho v_n}{2a} ~~~h<0,~~|h|~\mbox{small},
\label{fnuc1}
\end{equation}
which implies that $\Delta F\ge\Delta F_L$ as $h$ is negative in the
vapour. For small $h$ the variation in $\Delta F$ is linear ---
the exponent is 1.
This result holds for any system at fluid-fluid coexistence with dispersion
forces near a complete wetting transition.

Assuming Eq.~(\ref{rate}) holds for the rate
and using Eq.~(\ref{fnuc1}) for $\Delta F$ we see that if
$\Delta F_L(\mu,T)$ and $\sigma\tau^{-1}$ vary smoothly with through
$\mu$ around $\mu_c$ that near coexistence $\Delta F$ has the following
form
\begin{equation}
N_n=
\left\{
\begin{array}{ll}
N_{nL}+\left[N'_{nL}+N_{nL}\frac{A\delta\rho v_n}{2a}\right](\mu-\mu_c)
& ~~~~~ \mu<\mu_c   \\
N_{nL}+N'_{nL}(\mu-\mu_c) & ~~~~~ \mu>\mu_c \\
\end{array}\right. ,
\label{fnuc}
\end{equation}
where $N_{nL}$ is the nucleation rate in the liquid at coexistence,
$\mu=\mu_c$, and $N_{nL}'$ is the derivative of the nucleation rate,
with respect to the chemical
potential in the liquid, at $\mu=\mu_c$.
Equation (\ref{fnuc}) states that the
first derivative of the nucleation rate is discontinuous at the vapour-liquid
transition because this derivative contains
a contribution from the thickening wetting layer on the vapour side
of the coexistence curve but not on the liquid side. Figure
\ref{figcw} is a schematic of the variation of the rate
of heterogeneous nucleation near the
vapour-liquid transition (the solid curve).

\subsection{Critical wetting}

For complete wetting, the thickness $l$ of the
layer of liquid diverges as coexistence is approached, as in Eq.~(\ref{lcomp}).
Now, on moving away from the critical point of the vapour-liquid
transition, if the attraction of the surface for the molecules is
not too strong, then the surface-vapour interface may cease to be wet
by the liquid phase. Then as coexistence is approached the
thickness of the layer of liquid between the surface and
the vapour does not diverge, it remains finite. This is called
partial wetting \cite{schick}. The transition
from complete wetting to partial wetting is a phase transition.
It can be continuous, called a critical wetting transition, or
it can be first order. We will deal with each in turn.

First critical wetting. This occurs when the coefficient of the
$l^{-2}$ term in the free energy per unit area, $f$, changes sign.
To deal with this we need the next order term in an expansion
in $l^{-1}$. This is the $l^{-3}$ term and adding such a term
to the free energy of Eq.~(\ref{fcomp}) we have
\begin{equation}
f=\frac{t}{l^2}+\frac{b}{l^3}-hl,
\label{fcrit}
\end{equation}
where $t$ is a measure of the distance from the critical wetting
transition which occurs at $t=0$, and $b$ is a positive constant.
Above the transition, where there is complete wetting, as both
terms are of the same sign we can neglect the $l^{-3}$ as being small
for thick layers, then with $t$ fixed we recover Eq. (\ref{fcomp}).
The transition occurs at coexistence
so $h=0$. Then minimising $f$ we have
\begin{equation}
l=
\left\{
\begin{array}{ll}
\infty & ~~~~~ t\ge  0 \\
3b/(-2t) & ~~~~~ t<0,\\
\end{array}\right. 
\label{lcrit}
\end{equation}
$l$ varies as $(-t)^{-1}$ near and below the transition. The leading
order interaction of the nucleus with the vapour phase is still given
by the second term in Eq.~(\ref{fint}) when the layer is thick,
i.e., near the transition. Thus, using Eq.~(\ref{lcrit}) for $l$ we have
\begin{equation}
\Delta F(T)=
\left\{
\begin{array}{ll}
\Delta F_L(T)& ~~~ t\ge  0 \\
\Delta F_L(T)-\left[8Av_n/(27b^3)\right]t^3 & ~~~ t<0,\\
\end{array}\right. 
\label{fnuc_c}
\end{equation}
the difference between the nucleation barrier in the vapour
phase and that in the liquid varies as $t^3$ ($t$ small)
below the transition, they
are the same above it.
Again this holds for any system at fluid-fluid coexistence with dispersion
forces near a critical wetting transition. The exponent of three is
rather large, it means that $\Delta F$ and its first, second and
third derivatives are all continuous at the transition. The singularity
in $\Delta F$ is very weak and so detecting its effect on the
variation in the nucleation rate $N_n$ near the wetting transition
in an experiment may be very difficult.

Below the wetting transition, $t<0$, the surface is partially wet:
either covered by a thin film of molecules or with only a few molecules
on the surface.
Then as the coexistence curve
is crossed at constant $t<0$ we go from a nucleus on a surface which
is essentially in direct contact with the vapour (with at most a thin
film between them) to a nucleus on a surface in contact with the liquid
phase. The rate of heterogeneous nucleation then has a discontinuity
at coexistence, $\mu=\mu_c$. It will jump upwards, as shown schematically
in Fig.~\ref{figcw} (the dashed curve).

\subsection{First-order wetting}

Now for a first-order partial-wetting-to-complete-wetting
transition. This occurs when the $l^{-3}$ term is negative, $b<0$.
Then for stability we require a $l^{-4}$ term. This is just as in
a Landau expansion for the free energy near a phase transition.
Adding a $l^{-4}$ term with a positive coefficient $c$ to the $f$
of Eq.~(\ref{fcrit}) we have
\begin{equation}
f=\left(\frac{b^2}{4c}+t\right)\frac{1}{l^2}+\frac{b}{l^3}+\frac{c}{l^4}-hl,
\label{f1st}
\end{equation}
where in order to keep the transition at $t=0$ we have added a
constant, $b^2/(4c)$ to the coefficient of $l^{-2}$. At coexistence
$h=0$ and we minimise to obtain $l$
\begin{equation}
l=
\left\{
\begin{array}{ll}
\infty & ~~~ t\ge  0 \\
8c\left/\left[\left(b^2-32ct\right)^{1/2}-3b\right] \right.
& ~~~ t\le0,\\
\end{array}\right.
\label{l1st}
\end{equation}
giving a jump from $l=-2c/b$ to $\infty$ at the transition.
Putting this jump in Eq.~(\ref{fint}) for the free energy of
the nucleus we have a jump in the free energy barrier
of $Av_n/(-2c/b)^3$. Near the transition the barrier varies as
\begin{equation}
\Delta F(T)=
\left\{
\begin{array}{ll}
\Delta F_L(T)& ~~~ t\ge  0 \\
\Delta F_L(T)+\left[Av_n/(-2c/b)^3\right]\left[1-12ct/b^2+O(t^2)\right]
& ~~~ t<0,\\
\end{array}\right. 
\label{fnuc_1st}
\end{equation}
Above the transition, $t>0$, the free energy barriers are the same in the
vapour and liquid phases while just below the transition the
difference between the two is $Av_n/(-2c/b)^3$. The jump
in the free-energy barrier to heterogeneous nucleation will cause
a jump in the nucleation rate $N_n$, from Eq. (\ref{rate}).
As coexisting vapour and liquid phases are cooled the rate of
heterogeneous nucleation at surfaces in contact with the vapour
phase will jump downwards when the first-order wetting transition is
crossed. Assuming $\sigma$ and $\tau$ vary smoothly through the
transition the ratio of the nucleation rate $N_n$ just above the
wetting transition to that just below it is $\exp[Av_n/(-2c/b)^3]$.

In general we expect that the deeper we are into the region
where the crystal is the equilibrium phase, the larger is
the nucleation rate. Thus
we expect that if we cool coexisting vapour and liquid phases,
that the rate of heterogeneous nucleation will increase in
both, see Fig.~\ref{mut}.
However, we have just shown that if there is a first-order
wetting transition that the rate of heterogeneous nucleation will
jump {\em downwards} as we cross this transition. Potentially
at least, the nucleation rate may not be a monotonic function
of temperature: in the vapour phase
it may increase and then jump downwards as the
transition is crossed. A non monotonic variation in
$\Delta F$ is very rare, we are aware of only one example \cite{auer01}.
The non-monotonic variation in $\Delta F$ found by Auer and Frenkel
\cite{auer01}
is (presumably) not due the presence of another phase transition.

When the wetting transition at coexistence is first order, a prewetting
transition branches off from the coexistence curve into the vapour
off coexistence \cite{schick,bonn01}. This prewetting transition does not go
far into the vapour phase, it ends at a critical point which is at
a value of $h$ which depends on $b$ and $c$ but is always small. At the
prewetting transition there is a jump in the value of $l$, this jump
decreases as $h$ decreases until the jump reaches zero at the
prewetting critical point. At the prewetting transition as $l$ jumps then
so does the barrier to nucleation, from Eq.~(\ref{fint}). This
transition including the critical point may be calculated from the
free energy Eq.~(\ref{f1st}). As this free energy is analytic it yields
mean-field exponents for the critical point \cite{chaikin}.
Thus at the temperature of the prewetting critical point,
$t_{cp}$, and near the critical point the thickness difference
$l-l_{cp}\sim \mbox{sgn}(h-h_{cp})|h-h_{cp}|^{1/3}$, where $l_{cp}$ and
$h_{cp}$ are the liquid layer thickness and value of $h$ at the critical
point. This corresponds to the critical exponent $\delta=3$ -- its
mean-field value. Putting this variation of $l$ into our expression
for the interaction of the nucleus with the vapour phase across the
liquid layer, we obtain
\begin{equation}
\Delta F-\Delta F_{cp}\sim -\mbox{sgn}(h-h_{cp})|h-h_{cp}|^{1/3}
~~~~\mbox{mean-field},
\label{prewetdf}
\end{equation}
the nucleation barrier varies with the chemical potential minus that
at the critical point to the one third power. $\Delta F_{cp}$ is the
free energy of the nucleus at the prewetting critical point, and
Eq.~(\ref{prewetdf}) holds for $t=t_{cp}$ and $|h-h_{cp}|$ small.

Finally, we note that unlike complete and critical wetting,
a first-order wetting transition proceeds via
nucleation and growth, see Refs.~\cite{bonn01,bonn00}.
So for example on cooling
below the transition, a metastable thick wetting layer may persist,
where by metastable we mean the that the thickness of layer is not
the thickness which occurs at the absolute
minimum of the free energy Eq.~(\ref{f1st}).
Note that this layer is then doubly metastable, its free energy is
higher than that of a thinner film and the system with either of
these two layer thicknesses has of course a higher free energy than at true
equilibrium where there is dilute-fluid--crystal coexistence. Above
we have, for simplicity,
neglected the time taken to reach equilibrium thickness and
assumed that the thickness of the layer is always that at the minimum
in the free energy Eq.~(\ref{f1st}). In reality the transition
from a thick to thin layer will follow some set of dynamics which
will complicate the analysis. See Refs.~\cite{bonn00,bonn01} and
references therein for the dynamics at first-order wetting
transitions.
See Ref.~\cite{zohreh} for an explicit study of the analogous problem
of homogeneous nucleation near another transition where nucleation
of the equilibrium and a metastable phase compete.

\section{Comparison with homogeneous nucleation near a bulk critical
point}
\label{homo}

Essentially by definition, phase transitions are where the thermodynamic
functions of an equilibrium system have singularities. In earlier
work \cite{searxxx,sear01a,sear01b,zohreh}
we showed that the rate of {\em homogeneous} nucleation has a singularity
at a Ising-type phase transition in the {\em bulk}
and here we have shown that the rate
of {\em heterogeneous} nucleation has a singularity at
{\em surface} phase transitions
associated with an Ising-type phase transition in the bulk. The presence of a
singularity at a phase transition is common to both homogeneous and
heterogeneous nucleation.

Our findings here for heterogeneous nucleation near
a prewetting critical point and our earlier
findings for homogeneous nucleation are particularly closely related.
In both
cases we have a nucleus which is a small (point-like) perturbation
which couples to the order parameter of the transition \cite{sear01a,searxxx}.
The order parameter is the density for the bulk transition and
the thickness in the prewetting surface transition.
In earlier work \cite{searxxx}
we used scaling arguments to obtain the correct exponents for
homogeneous nucleation near the critical point of an Ising-type
transition. In
heterogeneous nucleation the nucleus
will couple to the order parameter
of the prewetting transition, which is an Ising-type transition in
two dimensions.
In Ref.~\cite{searxxx} we showed that near an Ising-type critical point
the behaviour is fixed and universal providing only that the
nucleus couples to the order parameter.
Thus heterogeneous nucleation
near a prewetting critical point is completely analogous to
homogeneous nucleation in a two-dimensional system near a bulk critical
point.
We can apply
the scaling approach of Ref.~\cite{searxxx}
to heterogeneous nucleation near a prewetting critical point.
We then obtain the
correct and universal exponents for the variation
of the free-energy barrier near the critical point.
For example along the prewetting
critical isotherm the free-energy barrier scales with distance
to the critical point as
\begin{equation}
\Delta F-\Delta F_{cp}\sim -\mbox{sgn}(h-h_{cp})|h-h_{cp}|^{1/15}
~~~~\mbox{universal},
\label{uni}
\end{equation}
where the exponent, which is $1/\delta$, is obtained from the
exact value $\delta=15$ for the Ising model in two dimensions. In
two dimensions the mean-field predictions for the critical exponents
like $\delta$ are very poor: the mean-field prediction, Eq.~(\ref{prewetdf}),
has an exponent which is five times too large \cite{note2}.
See Ref.~\cite{chaikin} for definitions of the critical exponents.
Equation (\ref{uni}) is just
Eq. (14) of Ref.~\cite{searxxx} in two dimensions. See that
reference for a derivation.

\section{Conclusion}

Almost invariably, a first-order phase transformation starts with
heterogeneous nucleation. The nucleus of the
new phase forms at a surface, see Fig.~\ref{fighet}. Thus,
the free energy barrier to the formation of the nucleus and therefore
the rate of nucleation in a phase depend sensitively on anything which happens
at the interface between the surface and the phase. If the phase
in contact with the surface is a vapour phase
close to a second, vapour-liquid,
phase transition, then if the surface attracts the molecules a
wetting layer may form at the surface. This is a layer of liquid at the
surface, separating the surface from the vapour. The wetting
layer will reduce the nucleation barrier greatly if the interfacial
tension between the liquid and the nucleus, $\gamma_{xl}$
is lower than that between the vapour and the nucleus, $\gamma_{xv}$.
We estimated in section \ref{hetero} that the reduction
in the surface contribution to the free energy barrier is of order
$10kT$. A large reduction which should be easily large enough
to observe in an experiment.
The formation of a wetting layer, either
as coexistence is approached (complete wetting), or along the
coexistence curve (critical wetting or a first-order wetting
transition) is a phase transition. We found that at complete wetting
the derivative of the barrier as a function of chemical potential
was discontinuous while the barrier itself has a discontinuity
as the temperature is varied through
a first-order wetting transition. Thus the rate of heterogeneous
nucleation has a discontinuity in its slope as the coexistence
curve is crossed at constant temperature, when there is complete wetting.
The rate of change of the nucleation rate in the vapour phase
just below coexistence is not the same as its rate of change in the
liquid phase just above coexistence. Above and below mean at values of the
chemical potential above and below that at coexistence. The
rate of heterogeneous nucleation has a discontinuity as a first-order
wetting transition is crossed.

Our model system is highly idealised, the surface is assumed perfectly smooth
and homogeneous, and the dynamics of the formation of wetting layers
have been neglected: the thickness was always taken to be at
equilibrium. Future work should address how the dynamics of formation
of wetting layers can effect nucleation; near a first-order
wetting transition this will presumably be analogous to homogeneous
nucleation near a metastable first-order bulk transition
\cite{zohreh}. Also, an understanding of
the effects of chemical heterogeneity and of curvature would be useful
as in practice surfaces will not be perfectly homogeneous or smooth and 
heterogeneous nucleation can occur on particles whose surfaces
are inherently curved. But the most urgent requirement is for
experiments on heterogeneous nucleation on simple, well characterised
surfaces.

It is a pleasure to acknowledge helpful discussions with A. Parry.
Work supported by EPSRC (GR/N36981).


\begin{thebibliography}{99}
\bibitem{xtalwet} Formation of the crystalline phase at the surface
does not have to be an activated process. If the crystal itself wets
the fluid-surface interface (this is a distinct phenomenon from the
wetting at vapour-liquid coexistence of the vapour-surface interface
by the liquid) then as the fluid-crystal transition is approached,
a layer of the crystal forms
at the surface and this can grow into a macroscopic crystal.
For example, if we move along the path of the arrow in Fig.~\ref{mut},
then as $\mu_X$ is approached from below, a layer of crystal forms
at the surface and at $\mu=\mu_X$ there is a macroscopic crystal formed.
Then the crystalline phase forms at $\mu=\mu_X$ and there is no barrier to its
formation. Throughout this work we are assuming that the crystalline
phase does not wet the fluid-surface interface.
\bibitem{note2} A mean-field theory \cite{sear01a,sear01b}
for the variation of the barrier to homogeneous
nucleation along the critical isotherm of a bulk critical point
yields the same exponent as Eq.~(\ref{prewetdf}).
The exponent for the variation in the nucleation
barrier along the critical isotherm was not calculated explicitly
in either of Refs.~\cite{sear01a,sear01b} but the exponent is
implicit in Eq.~(14) of Ref.~\cite{sear01b}. This equation contains
a factor of $(m_c-m)^2$ which dominates the
variation along the critical isotherm near the critical point.
$m$ is the order parameter of the
transition (0 at the critical point) and $m_c$ is the value of the
order parameter in the core of the nucleus, $m_c=O(1)$. Near the
critical point $m$ is small and we may expand to obtain
$m_c^2-2m_cm+O(m^2)$. Near the
critical point the second term scales as the order parameter which
scales as
$\mbox{sgn}(h-h_{cp})|h-h_{cp}|^{1/3}$
within a mean-field theory. The first term is just a constant.
\end{thebibliography}
\end{document}